\documentclass[aps,twocolumn,superscriptaddress,showpacs]{revtex4}
\usepackage{graphicx}
\begin{document}
\title{Examining exotic structure of proton-rich nucleus $^{23}$Al}

\author{D. Q. Fang\footnote{Corresponding author. Email: dqfang@sinap.ac.cn}}
\author{W. Guo}
\author{C. W. Ma}
\author{K. Wang}
\author{T. Z. Yan}
\author{Y. G. Ma}
\author{X. Z. Cai}
\author{W. Q. Shen}
\affiliation{Shanghai Institute of Applied Physics, Chinese Academy
of Sciences, Shanghai 201800, People's Republic of China}

\author{Z. Z. Ren}
\affiliation{Department of Physics, Nanjing University, Nanjing 210008, People's Republic of China}
\author{Z. Y. Sun}
\affiliation{Institute of Modern Physics, Chinese Academy of
Sciences, Lanzhou 730000, People's Republic of China}

\author{J. G. Chen}
\author{W. D. Tian}
\author{C. Zhong}
\affiliation{Shanghai Institute of Applied Physics, Chinese Academy
of Sciences, Shanghai 201800, People's Republic of China}

\author{M. Hosoi}
\affiliation{Department of Physics, Saitama University, Saitama 338-8570, Japan}

\author{T. Izumikawa}
\affiliation{Department of Physics, Niigata University, Niigata 950-2181, Japan}

\author{R. Kanungo}
\affiliation{TRIUMF, 4004 Wesbrook Mal, Vancouver, British Columbia, V6T 2A3, Canada}
\author{S. Nakajima}
\affiliation{Department of Physics, Saitama University, Saitama 338-8570, Japan}
\author{T. Ohnishi}
\affiliation{Institute of Physical and Chemical Research (RIKEN),
Wako-shi, Saitama 351-0198, Japan}
\author{T. Ohtsubo}
\affiliation{Department of Physics, Niigata University, Niigata 950-2181, Japan}
\author{A. Ozawa}
\affiliation{Institute of Physics, University of Tsukuba, Ibaraki 305-8571, Japan}
\author{T. Suda}
\affiliation{Institute of Physical and Chemical Research (RIKEN),
Wako-shi, Saitama 351-0198, Japan}
\author{K. Sugawara}
\author{T. Suzuki}
\affiliation{Department of Physics, Saitama University, Saitama 338-8570, Japan}
\author{A. Takisawa}
\affiliation{Department of Physics, Niigata University, Niigata 950-2181, Japan}
\author{K. Tanaka}
\affiliation{Institute of Physical and Chemical Research (RIKEN),
Wako-shi, Saitama 351-0198, Japan}
\author{T. Yamaguchi}
\affiliation{Department of Physics, Saitama University, Saitama 338-8570, Japan}
\author{I. Tanihata}
\affiliation{TRIUMF, 4004 Wesbrook Mal, Vancouver, British Columbia, V6T 2A3, Canada}

\date{\today}

\begin{abstract}
The longitudinal momentum distribution ($P_{//}$) of fragments after
one-proton removal from $^{23}$Al and reaction cross sections ($\sigma_R$)
for $^{23,24}$Al on carbon target at $74A$~MeV have been measured.
The $^{23,24}$Al ions were produced through projectile fragmentation of
135$A$~MeV $^{28}$Si primary beam
using RIPS fragment separator at RIKEN.
$P_{//}$ is measured by a direct time-of-flight (TOF) technique,
while $\sigma_R$ is determined using a transmission method.
An enhancement in $\sigma_R$ is observed for $^{23}$Al
compared with $^{24}$Al. The $P_{//}$ for $^{22}$Mg fragments
from $^{23}$Al breakup has been obtained for the first time.
FWHM of the distributions has been determined to be 232$\pm$28~MeV/c.
The experimental data are discussed by using Few-Body Glauber model.
Analysis of $P_{//}$ demonstrates a dominant $d$-wave configuration
for the valence proton in ground state of $^{23}$Al, indicating that
$^{23}$Al is not a proton halo nucleus.
\end{abstract}

\maketitle

Since the pioneering measurements of
interaction cross sections ($\sigma_I$) and observation of a
remarkably large $\sigma_I$ for $^{11}$Li~\cite{TAN85PRL,TAN92PLB},
exotic structures like neutron halo or skin in light neutron-rich
nuclei have been found. Experimental measurements of reaction
cross section ($\sigma_R$), fragment momentum distribution
($P_{//}$) after one or two nucleons removal, quadrupole moment and
Coulomb dissociation have been demonstrated to be very effective
methods to investigate nuclear halo structure.
The neutron skin or halo nuclei $^{6,8}$He, $^{11}$Li, $^{11}$Be,
$^{19}$C etc.~\cite{TAN85PRL,TAN85PLB,TAN92PLB,FUK91,BAZ,NAK,OZA01},
have been identified by these experimental methods.
Due to Coulomb barrier, identification of
a proton halo is more difficult
compared to a neutron halo. The quadrupole moment, $P_{//}$ and
$\sigma_R$ data indicate a proton halo in
$^{8}$B~\cite{MIN,SCH,WAR1,NEG,FUK}, whereas no enhancement is
observed in the measured $\sigma_I$ at relativistic energies~\cite{OBU}.
The proton halo in $^{26,27}$P
and $^{27}$S has been predicted theoretically~\cite{BRO,REN}.
Measurements of $P_{//}$ have shown a proton halo character in
$^{26,27,28}$P~\cite{NAV}.

\begin{figure*}[t]
\centering
\begin{minipage}{10.5cm}
\includegraphics[height=10.5cm,angle=-90.]{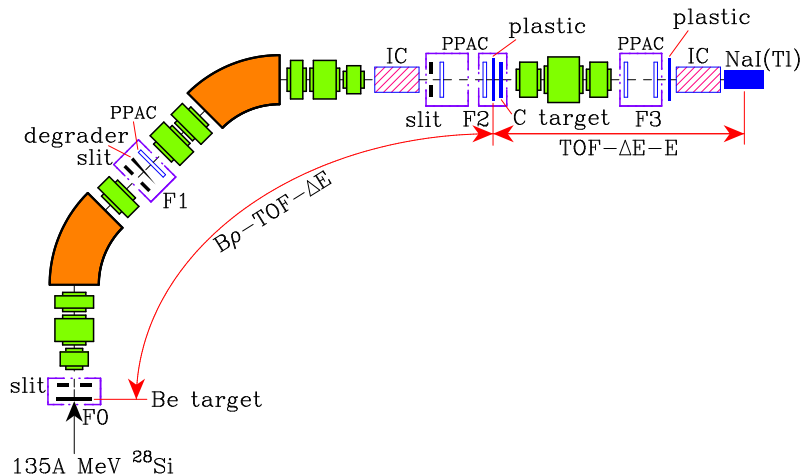}
\end{minipage}\hskip 1.cm%
\begin{minipage}{4.cm}
\caption{(Color online) Experimental setup at the fragment separator RIPS.}
\label{rips-setup}
\end{minipage}
\end{figure*}

Proton-rich nucleus $^{23}$Al has a very small separation energy
($S_p=0.125$~MeV)~\cite{AUD93} and is a possible candidate of
proton halo. An enhanced $\sigma_R$ for $^{23}$Al has
been observed in a previous experiment~\cite{CAI02,ZHA}.
To reproduce the $\sigma_R$ for $^{23}$Al
within framework of Glauber model, a dominating
$2s_{1/2}$ component for the valence proton
is shown~\cite{CAI02}. A long tail in proton
density distribution has been extracted for $^{23}$Al
which indicated halo structure.
The spin and parity ($J^{\pi}$) for ground state of $^{23}$Al
has been deduced to be 5/2$^+$ in a recent measurement of magnetic
moment~\cite{OZA06}. This result favors a $d$-wave configuration
for  the valence proton in $^{23}$Al.
But it does not eliminate the possibility of a $s$-wave valence
proton if its $^{22}$Mg core is in excited state.
Therefore it will be very important to determine configuration
of the valence proton for $^{23}$Al.
As we know, $P_{//}$ of the fragment
carries structure information of the projectile.
However, there are no such
experimental data for $^{23}$Al up to now.
In this paper we will report simultaneously measurements of
$\sigma_R$ and $P_{//}$ for $^{23}$Al and also $\sigma_R$
for $^{24}$Al.

The experiment was performed at the RIken Projectile fragment
Separator (RIPS) in RIKEN Ring Cyclotron Facility.
The experimental setup is shown in Fig.~\ref{rips-setup}.
Secondary beams were generated by fragmentation reaction
of 135$A$~MeV $^{28}$Si primary beam on a $^9$Be production
target in F0 chamber.
In the dispersive focus plane F1, an Al wedge-shaped degrader
(central thickness; 583.1 mg/cm$^2$, angle; 6 mrad) was installed.
A delay-line readout Parallel Plate Avalanche
Counter (PPAC) was placed to measure the beam position.
Then the secondary beam was directed onto the achromatic focus F2.
Two delay-line readout PPACs were installed to determine the beam position
and angle. An ion chamber (200$\phi\times780$mm) was used to measure
energy loss ($\Delta E$) of the secondary beams~\cite{KIM05}. An ultra-fast
plastic scintillator (0.5 mm thick) was placed before a carbon reaction
target (377 mg/cm$^2$ thick) to measure time-of-flight (TOF) from
the PPAC at F1. The particle identification before the reaction target
was done by means of $B\rho-\Delta E-$TOF method.

\begin{figure}[b]
\centering\includegraphics[width=8.cm]{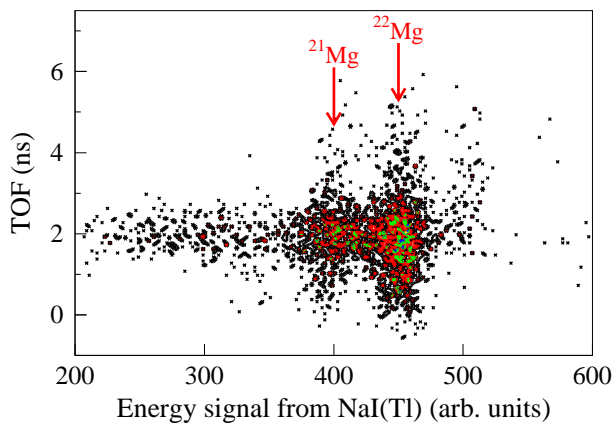}
\caption{(Color online) Particle identification at F3 by the bidimensional
plot between TOF from F2 to F3 and energy signal from NaI(Tl)
(corrected with TOF).
}
\label{pid}
\end{figure}

After the reaction target, a quadrupole triplet was used to transport and
focus the beam onto F3 ($\sim$ 6 m from F2). Two delay-line readout PPACs
were used to monitor the beam size and emittance angle. Another plastic
scintillator (1.5~mm thick) gave a stop signal of the TOF from F2 to F3.
A smaller ion chamber (90$\phi\times650$mm) was used to measure energy
loss ($\Delta E$) of the beam. Total energy ($E$) was measured by a NaI(Tl)
detector. The particles were identified by TOF$-\Delta E$$-$$E$ method.
An example of typical particle identification spectra at F3 for
the fragment from $^{23}$Al breakup
is shown in Fig.~\ref{pid}. In this spectrum, fragments with different
nuclear charge were already subtracted by TOF and $\Delta E$ method.

Under assumption of a sudden valence-nucleon removal, the momentum
distribution of fragments can be used to describe that of the valence
proton.
The $P_{//}$ of fragments from breakup reactions was determined from
the TOF between the two plastic scintillators installed at F2 and F3.
Position information measured by the PPAC at F1 was used to derive
incident momentum of the beam. The momentum of fragment relative to
the incident projectile in laboratory frame was transformed into that
in the projectile rest frame using Lorentz transformation.

\begin{figure}[t]
\centering\includegraphics[width=7.cm]{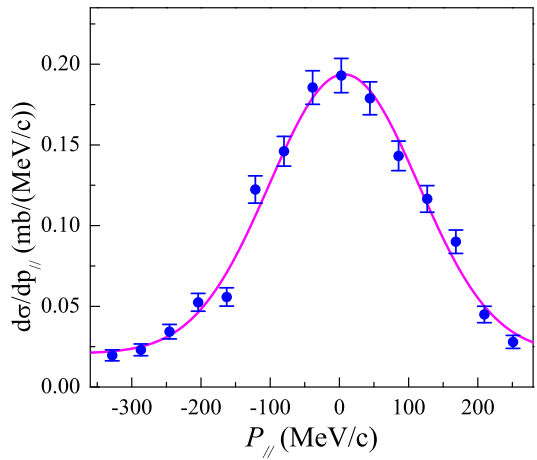}
\caption{(Color online) $P_{//}$ distribution of fragment $^{22}$Mg
after one-proton removal from $^{23}$Al. The closed circles with
error bars are the present experimental data, the solid curve is
a Gaussian fit to the data.}
\label{PP-23Al}
\end{figure}

In order to estimate and subtract reactions of the projectile
in material other than the carbon target, measurements without
the reaction target were also performed and the beam energy was
reduced by an amount corresponding to the energy loss in the target.
For one-proton removal reactions of $^{23}$Al,
this background was carefully reconstructed and subtracted based on
ratio of fragments to unreacted projectile identified in the
target-out measurement and also broadening effect of the carbon target
on $P_{//}$. The obtained momentum distribution of $^{22}$Mg
fragments from $^{23}$Al breakup in the carbon target
at 74$A$~MeV is shown in Fig.~\ref{PP-23Al}.
We normalized experimental counts to the measured
one-proton removal cross section ($\sigma_{-1p}$)
so that $\sum N(p_{\mbox{i}})\Delta p_{||}$ equals
$\sigma_{-1p}$.
A Gaussian function was used to fit the distributions.
The full width at half maximum (FWHM) was determined to be
232$\pm$28~MeV/c after unfolding the Gaussian-shaped
system resolution (41 MeV/c).
The FWHM is consistent with Goldhaber model's prediction
(FWHM=212~MeV/c with $\sigma_0=90$~MeV/c) within the error bar~\cite{GOL}.
Since magnetic fields of the quadruples between F2 and F3 were optimized
for the projectile in the measurement,
momentum dependence of transmission from F2 to F3 for fragments
was simulated by the code MOCADI~\cite{IWA}. The effect of transmission
on the width of $P_{//}$ distribution was found to be negligible which
is similar with the conclusion for neutron-rich nuclei~\cite{YAM,FAN04}.
Using the estimated transmission value, the one-proton removal cross
sections for $^{23}$Al was obtained to be 63$\pm$9 mb.

Reaction cross section is determined using the transmission method:
\begin{equation}
\sigma_R=\frac{1}{t} \ln\left(\frac{\gamma_0}{\gamma}\right)
\label{Transmit}
\end{equation}
where $\gamma$ and $\gamma_0$ denote ratio of unreacted outgoing and
incident projectiles for target-in and target-out cases, respectively;
$t$ thickness of the reaction target, i.e., number of particle
per unit area.

The $\sigma_R$ of $^{23,24}$Al at 74$A$~MeV were obtained to be
1609$\pm$79 mb and 1527$\pm$60 mb, respectively.
The errors include statistical and systematic uncertainties.
Probability of inelastic scattering reaction was estimated
to be very small ($<1\%$), e.g., the inelastic cross section is
only around 11~mb for $^{23}$Al which is much smaller than the
error of $\sigma_R$.

\begin{figure}[t]
\centering\includegraphics[width=8.cm]{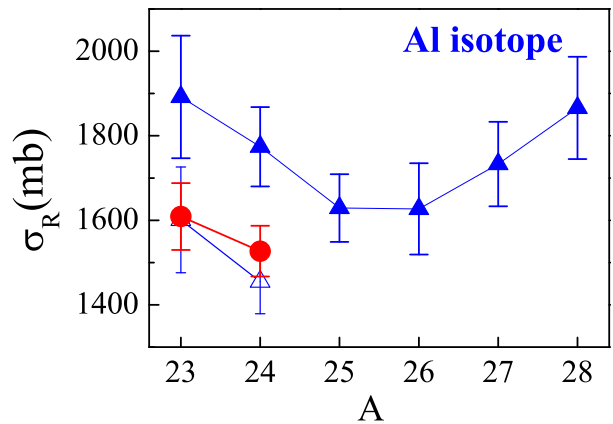}
\caption{(Color online) The mass dependence of $\sigma_R$ for Al isotopes.
The solid circles are results of the present experiment ($E=74A$~MeV),
the solid triangles are the previous experimental data
($E\sim 30A$~MeV)~\cite{CAI02}, and the open
triangles are the previous data scaled to $74A$~MeV.}
\label{sigma}
\end{figure}

Results of previous and current experiments are shown in Fig.~\ref{sigma}.
Since the energy is different in two experiments, the previous $\sigma_R$
data at $\sim 30A$~MeV~\cite{CAI02} were scaled to the present energy
($74A$~MeV) using a phenomenological formula~\cite{SHE}.
First the radius parameter ($r_0$) in this formula was adjusted to
reproduce the $\sigma_R$ at $\sim 30A$~MeV, then the same $r_0$ was
used to calculate the $\sigma_R$ at $74A$~MeV.
As shown in Fig.~\ref{sigma}, the $\sigma_R$ of $^{23,24}$Al from
present and previous experiments are in good agreement.
And we observed a small enhancement in $\sigma_R$ for $^{23}$Al
in our data again.

To interpret the measured reaction cross sections and momentum distributions,
we performed a Few-Body Glauber model (FBGM) analysis for $P_{//}$ of
$^{23}$Al~$\rightarrow$~$^{22}$Mg processes and $\sigma_R$ of
$^{23,24}$Al~\cite{OGA92,OGA94,IBR}.
In this model, a core plus proton structure is assumed for the projectile.
The total wavefunction of the nucleus is expressed as
 \begin{equation}
\Psi=\sum_{ij}\psi^{i}_{\mbox{core}}\phi^{j}_{0},
\end{equation}
where $\psi_{\mbox{core}}$ and $\phi_{0}$ are wavefunctions of
the core and valence proton; $i$,$j$ denote different
configurations for the core nucleus and valence proton, respectively.
For the core, harmonic oscillator (HO) functions were used for the
density distributions. The wavefunction of the
valence proton was calculated by solving the eigenvalue problem in
a Woods-Saxon potential.
The separation energy of the last proton is reproduced by adjusting the
potential depth. In the calculation, the diffuseness and radius parameter
were chosen to be 0.67~fm and 1.27~fm, respectively~\cite{YAM}.

In the recent $g$-factor measurement using a $\beta$-NMR method,
the spin and parity for ground state of $^{23}$Al
is shown to be $5/2^+$. It gives a strong restriction on the
possible structure of this nucleus. Assuming $^{22}$Mg$+p$
structure, three most probable configurations for
$J^{\pi}=5/2^+$ of $^{23}$Al are: $0^+$$\otimes$$1d_{5/2}$,
$2^+$$\otimes$$1d_{5/2}$ and $2^+$$\otimes$$2s_{1/2}$~\cite{OZA06}.
The $s$-wave configuration is therefore possible for the core
in the excited state.

\begin{figure}[t]
\centering\includegraphics[width=7.3cm]{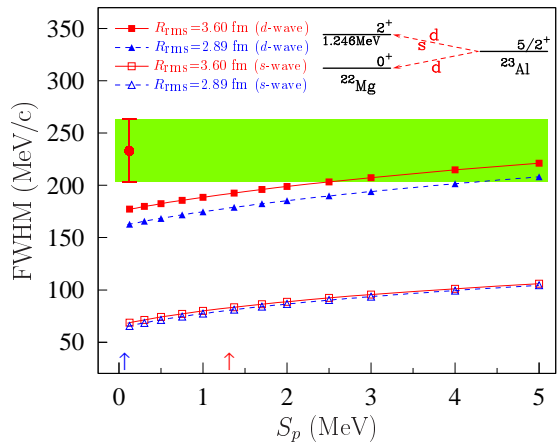}
\caption{(Color online) The dependence of FWHM for the $P_{//}$
distribution after one-proton removal of $^{23}$Al on the separation
energy of the valence proton. The solid circles with error bars is
result of the present experiment, the shaded area refers to error
of the data. The solid and open squares are the FBGM calculations
for the $d$ and $s$-wave configuration of the valence proton with
the core $R_{\mbox{rms}}=3.6$~fm. The solid and open triangles are
for the core $R_{\mbox{rms}}=2.89$~fm. The lines are just for
guiding the eyes. The two arrows refer to the separation energy
of 0.125~MeV and 1.37~MeV (the excitation energy for the first
excited state of $^{22}$Mg plus the experimental one proton
separation energy of $^{23}$Al).}
\label{fwhm}
\end{figure}

The momentum distributions for the valence proton in
$s$ or $d$-wave configuration are calculated by using FBGM.
In this calculation, the parameters
$\alpha$ and $\sigma_{NN}$ in the profile function
$\Gamma(b) =\frac{1-i\alpha}{4\pi\beta^2}\sigma_{NN}
\exp(-\frac{b}{2\beta^2})$ ($b$ is the impact parameter)
are taken from Ref.~\cite{OGA94}.
The range parameter ($\beta$) is calculated by the formula
which is determined by fitting the $\sigma_{R}$ of $^{12}$C + $^{12}$C
from low to relativistic energies~\cite{ZHE}.
$\beta$ is 0.35~fm at $74A$~MeV.
To fix the core size, the width parameters in the HO density
distribution of $^{22}$Mg were adjusted to reproduce the
experimental $\sigma_I$ data at around $1A$~GeV~\cite{SUZ}.
The extracted effective root-mean-square matter radius
($R_{\mbox{rms}}\equiv<r^2>^{1/2}$) for $^{22}$Mg
is $2.89\pm0.09$ fm.
To see the separation energy dependence, the FWHM of $P_{//}$
is determined assuming an arbitrary separation energy
in calculation of the wavefunction for the valence proton
in $^{23}$Al and shown in Fig.~\ref{fwhm}.
If we adopt a larger radius of $R_{\mbox{rms}}=3.6$ fm
for $^{22}$Mg to see the core size effect on $P_{//}$,
we obtained solid and open squares of FWHM in Fig.~\ref{fwhm}.
The one proton separation energies for $^{22}$Mg in the ground
and excited ($J^{\pi}=2^+$, $E_x=1.25$~MeV) states are taken as
0.125~MeV and 1.375~MeV ($E_x+0.125$~MeV). Those two values are marked
by two arrows in Fig.~\ref{fwhm}.
In this figure, we can see that the width for the $s$ and $d$-wave
are obviously separated.
The width for the $s$-wave is much lower than the experimental data,
while that of the $d$-wave is close to the experimental FWHM.
With the increase of $S_p$, the width of $P_{//}$ increases slowly.
It means that $P_{//}$ will become wider for $^{22}$Mg in the excited
state. The effect of the core size on $P_{//}$ is negligible for
the $s$-wave and small for the $d$-wave configuration.
The larger sized core will give a little wider $P_{//}$ distribution.
From comparison of the FBGM calculation with
the experimental data in Fig.~\ref{fwhm}, it is clearly shown
that the valence proton in $^{23}$Al is dominantly in the $d$-wave
configuration. The possibility for the $s$-wave should be very small.
Furthermore, it is possible to have an excited core inside $^{23}$Al.
This is consistent with the shell model calculations and also the
Coulomb dissociation measurement~\cite{OZA06,GOM05}.

\begin{figure}[t]
\centering\includegraphics[width=8.cm]{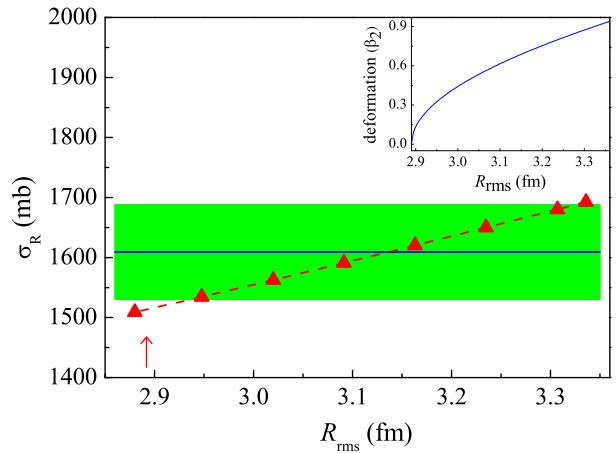}
\caption{(Color online) The dependence of $\sigma_R$ at $74A$~Mev on
the core size ($R_{\mbox{rms}}$). The horizontal line is the
experimental $\sigma_R$ value, the shadowed area is the error of
$\sigma_R$. The triangles denote the FBGM calculations.
The size of $^{22}$Mg obtained by fitting the $\sigma_I$ data
at around $1A$~GeV is marked by an arrow. The inset shows
the relationship between the quadrupole deformation parameter
($\beta_2$) and size of the core, for details see the text.
}
\label{core-size}
\end{figure}

From above discussions of $P_{//}$, the valence proton in $^{23}$Al
is determined to be in the $d$-wave configuration, which is used in
the following calculations. In the calculation of $\sigma_R$ for $^{23}$Al
using the FBGM, at first $R_{\mbox{rms}}=2.89\pm0.09$ fm is used for
its $^{22}$Mg core by reproducing the $\sigma_I$ data as described above.
But the calculated $\sigma_R$ is much lower than the obtained
$\sigma_R$ data.
One reason may be due to the global underestimation of $\sigma_R$
found at intermediate energies in the Glauber model~\cite{OZA2}.
Different method has been tried to correct this problem~\cite{FUK91,ZHE,TAK}.
These corrections are performed for almost light stable nuclei.
The $\sigma_{R}$ of $^{24}$Al is calculated with the size of
its $^{23}$Mg core determined by fitting $\sigma_I$ at around $1A$~GeV~\cite{SUZ}.
But the calculated $\sigma_R$ for $^{24}$Al is only 1430 mb which
is $\sim10\%$ lower than the present data.
It was shown that scope of the discrepancy between the Glauber model
calculation and experimental data is large even for stable nuclei~\cite{OZA2}.
To correct the possible underestimation for nuclei with $A>20$,
we adjusted the range parameter to fit the $\sigma_R$
of $^{24}$Al from the present measurement.
And $\beta=0.8$~fm is obtained when the $\sigma_R$ of $^{24}$Al at $74A$ MeV
is reproduced.
Using this range parameter, the calculated $\sigma_R$ value of $^{23}$Al
is still smaller than the data.
Similar puzzle is also encountered for some neutron-rich
nuclei. The large $\sigma_I$ cannot be reproduced by the FBGM even for
the valence neutron in the $s$-wave for $^{19}$C and $^{23}$O.
One way is to enlarge the core size to reproduce the experimental
$\sigma_R$~\cite{KAN00,KAN02}.
Here we changed the core size by adjusting width parameters
in the HO density distribution of $^{22}$Mg.
The dependence of $\sigma_R$ for $^{23}$Al on the core size is
calculated and shown in Fig.~\ref{core-size}.
The calculated results indicate that the core size is
$3.13\pm0.18$~fm when the experimental $\sigma_R$ data of $^{23}$Al
is reproduced (8$\pm$7\% larger than the size of the bare $^{22}$Mg nucleus).

In order to reproduce the $\sigma_R$ of $^{23}$Al from the current work,
a larger sized core is deduced within the framework of the spherical Glauber
model. It should be pointed out that this enlarged core may not necessarily
reflect increased physical size of the nucleus. The negligence of some
specific effects in the Glauber model could lead to the larger sized core.
The possible reasons for the enlargement will be discussed qualitatively below.
The effect of quadrupole deformation
($\beta_2$, the parameter describing the deformation)
on the rms radius can be expressed as
$R_{\mbox{rms}}^{\beta_2}=\sqrt{(1+\frac{5}{4\pi}\beta_2^2)}
R_{\mbox{rms}}^{\beta_2=0}$~\cite{CHR}.
As shown in the inset of Fig.~\ref{core-size},
$R_{\mbox{rms}}$ of the core changes quickly with the increase of
$\beta_2$. This simple relationship between $R_{\mbox{rms}}$ and
$\beta_2$ indicates that a deformed core inside $^{23}$Al will give
a larger sized $^{22}$Mg.
In order to reproduce the $\sigma_R$ of $^{23}$Al, the lower limit
of $R_{\mbox{rms}}$ for the core is 2.95 fm as we can see from the
calculated results in the figure.
If we assume that the shape of $^{22}$Mg as a nucleus is spherical and
enlargement of the core is due to deformation,
the lower limit of $\beta_2=0.3$ for the core could be deduced from
the inset of Fig.~\ref{core-size}. Deformation of $\beta_2=0.6$
will give around 8$\%$ larger radius for the $^{22}$Mg core.
The experimental and theoretical investigations have demonstrated the
deformation for $^{22}$Mg. The experimental $\beta_2$ is 0.566~\cite{RAM},
the calculated $\beta_2$ by RMF and generalized hybrid derivative
coupling model are around 0.4 and 0.6, respectively~\cite{LAL,MIT}.
If the bare nucleus $^{22}$Mg is
deformed, the above analysis indicates that $^{22}$Mg as a core in $^{23}$Al
may have larger deformation as compared with $^{22}$Mg as a nucleus.
Additionally, the first excited state of $^{22}$Mg was calculated within
the RMF framework and its $R_{\mbox{rms}}$
is obtained to be around 2.4\%
larger than that of the ground state~\cite{REN2,CHE}.
Thus the core excitation
effect may also contribute to the larger size for $^{22}$Mg.
As demonstrated by the shell model calculations, the configuration
of $^{22}$Mg (ground state) plus a $d$-wave proton is dominant in
$^{23}$Al~\cite{OZA06}. If deformation and excitation effects
exist in the core, the first one may be the main component.

In summary, the longitudinal momentum distribution of fragments
after one-proton
removal for $^{23}$Al and reaction cross sections for $^{23,24}$Al
were measured. An enhancement was observed for the $\sigma_R$ of
$^{23}$Al. The $P_{//}$ distributions were found to be wide and
consistent with the Goldhaber model's prediction.
The experimental $P_{//}$ and $\sigma_R$ results were discussed within
framework of the Few-Body Glauber model.
We determined the valence proton to be a
dominant $d$-wave configuration in the ground state of $^{23}$Al.
It indicates no halo structure in this nucleus.
But a larger sized $^{22}$Mg core was deduced in order
to explain both the $\sigma_R$ and $P_{//}$ distributions
within framework of the spherical Few-Body Glauber model.
It is pointed out that deformation and core excitation effects may be
two main reasons for the extracted larger sized core.
Further theoretical investigations are needed to extract more specific
structure information for $^{23}$Al from the experimental data.

\section*{Acknowledgements}
The authors are very grateful to all of the staff at the RIKEN
accelerator for providing stable beams during the experiment.
The support and hospitality from the RIKEN-RIBS laboratory
are greatly appreciated by the Chinese collaborators.
This work was partially supported by the National Natural Science
Foundation of China (NNSFC) under Grant No. 10405032, 10535010,
10405033 and 10475108, Shanghai Development
Foundation for Science and Technology under contract No. 06QA14062,
06JC14082 and 05XD14021, the Major State Basic
Research Development Program in China under Contract No. 2007CB815004
and the Knowledge Innovation Project of Chinese Academy of Sciences
under Grant No. KJCX3.SYW.N2.

\end{document}